\documentclass{article}

\usepackage[english]{babel}
\usepackage[utf8]{inputenc}
\usepackage[T1]{fontenc}

\usepackage{url}
\usepackage{graphicx}
\usepackage{amsmath}
\usepackage[linesnumbered,ruled]{algorithm2e}
\usepackage{subfigure}

\begin{document}

\title{RALL --- Routing-Aware Of Path Length, Link Quality, And
  Traffic Load For Wireless Sensor Networks}

\author{
  Vinícius N. Medeiros, Douglas V. Santana, \\
  Bruno Silvestre, Vinicius da C. M. Borges \\
  {\textit {\{viniciusnunesmedeiros, douglassantana\}@inf.ufg.br}} \\
  {\textit {\{brunoos, vinicius\}@inf.ufg.br}} \\
  \\
  Instituto de Informática -- Universidade Federal de Goiás \\
  Goiânia, GO -- Brasil
}

\date{15 September 2016}

\maketitle

\begin{abstract}


Due to the enormous variety of application scenarios and ubiquity,
Internet of Things (IoT) brought a new perspective of applications for
the current and future Internet. The Wireless Sensor Networks provide
key devices for developing the IoT communication paradigm, such as the
sensors collecting various kind of information and the routing and MAC
protocols.  However, this type of network has strong power consumption
and transmission capacity restrictions (low speed wireless links and
subject to interference).  In this context, it is necessary to develop
solutions that enable a more efficient communication based on the
optimized utilization of the network resources.  This papers aims to
present a multi-objective routing algorithm, named Routing-Aware of
path Length, Link quality, and traffic Load (RALL), that seeks to
balance three objectives: to minimize bottlenecks, to minimize path
length, and to avoid links with low quality.  RALL results in good
performance when taking into consideration delivery rate, overhead,
delay, and power consumption.

\end{abstract}

\section{Introduction}


The Internet has been employed more and more, and becoming an
essential tool for humans. Nowadays, not only people use this
information and communication technology, as well as machines employ
it to communicate with each other, making measurement of various types
of information. The Internet of Things (IoT) paradigm allows these
machines to communicate and feel the environment~\cite{Atzori2010}.


Wireless Sensor Networks (WSNs) play an important role within the IoT
and have gained increasing prominence as part of ubiquitous computing
in various environments, such as industry, smart cities, smart spaces,
smart grids, health monitoring environmental, real-time multimedia
applications \cite{Atzori2010,Oliveira11,Alanazi2015}. WSNs consist of
various nodes that generally have strong processing, battery, and
memory constraints. Typically, each node has only one radio interface
(generally 802.15.4 CSMA/CA) with a fixed transmission rate (250
kbps)~\cite{Yick2008}.


Usually, a flat model is employed to organize the network, where all
the nodes play the same role in sensing, processing, and (re)broadcast
packets. The information is forwarded (routed) node by node until it
reaches the sink node, where it is
processed~\cite{Akyildiz2007,Niculescu2005,Radi2011}. This paper will
treat the routing optimization problem based on three objectives: (i)
minimize the path length (number of hops), (ii) minimize the
bottleneck, and (iii) minimize the interference in the path. At the
routing, each hop could increase the delay.



This article is organized as follows: In Section~\ref{sec:modelagem},
we present a model for the routing problem of multi-objective
optimization. In Section~\ref{sec:descricaoAlgoritmo}, we describe the
proposed algorithm to determine a solution for the model. In
Section~\ref{sec:resultados}, we present the simulation results. In
Section~\ref{sec:relatedwork}, we analyze the most relevant related
work on routing approaches for WSNs. The conclusion and future work is
carried out in Section~\ref {sec:conclusao}.

\section{System Model} \label{sec:modelagem}


WSNs consist of various sensors, distributed in an area, that enable
data collection. Most of these sensors monitor or interact with the
environment in which they are. A node in particular, called sink, is
designed to gather all the data collected by the sensors. The sink
node is attached to a machine with more processing capacity
(e.g. servers in cloud computing environment) so that collected data
can be transformed into information and knowledge, to be used by
various applications.


WSNs are modeled as a graph $G = (V, E)$, where $V$ is the set of
vertices that represents the sensors, $E$ is the set of communication
links between two network devices. The link $e_{sd} \in E$ between
$s,d \in V$ exists only if the device $s$ accomplish a data
transmission to the node $d$, i.e. the node $d$ must be in the
transmission range of $s$. Some metrics can be employed for
determining if a link between two sensors can actually be used. Every
sensor $s \in V - \{i\}$, where $i$ is the sink, is responsible for
the origin of a data stream named $f_s \in F$, where $F$ is defined as
the set of all flows generated in the network. All flows of WSNs have
the sink node $i$ as destination node.


The establishment of a path is demanded on the graph $G$ so that the
data flow can reach the sink node. The selection of each flow path
must follow some objectives to ensure certain network characteristics,
such as packet delivery rate, energy consumption, delay, etc. A larger
number of hops require the activation of more links, that generates
more use of the transmission medium and may lead to more contention or
interference as well as end-to-end delay~\cite{Flushing2013}.


Load balancing aims at distributing the flows over the network
uniformly. This prevents a small set of nodes to be used for the
majority of the routes, leading to a greater burden on those nodes. It
is important to point out that the nodes have limited resources, so
that load can cause, for example, the reduction of the network
lifetime, due to a higher consumption of battery to forward packets,
as well as for packet loss, given the small amount of memory in the
nodes for packet storage~\cite{Kang2004}.



The interference degrades the link quality~\cite{Flushing2013}. It can
be caused by the number of neighbors that are in communication range
of each node and/or other wireless networks that generally adopt
communication standards using same spectrum band. As nodes in a WSNs
use this shared band, and they employ only a radio interface and a
single communication channel, more nodes in the network result in
higher level of interference. This implies a greater chance of
collision (requiring packet retransmission) because there is a greater
probability that the channel is busy. As a result, interference
strongly impacts the effective transmission capacity. Thus, the routes
that pass through nodes with a greater degree of interference may
suffer more delay in delivery and/or more retransmission of packets,
as well as may result in larger battery consumption.


It can be noticed interrelationships between these three
objectives. The increase in the number of hops in the path is a common
consequence when a routing solution seeks to minimize
overload~\cite{Flushing2013}. However, increasing path length causes
more nodes activation, in consequence more interference. It is
important to notice that not always shorter paths generate less
interference in the WSN as a whole. For instance, the shortest paths
may suffer interference due to the existence of neighbors or other
wireless networks/devices that are in the same range area of these
paths.

Following we present the construction of an optimization model for the
route establishment on WSN with three objectives: minimizing the
number of hops, minimizing the network bottleneck on the links, and
minimizing the use of low quality links. These three objectives are
described and modeled in the following paragraphs.

\paragraph{Number of Hops}


Minimizing the number of hops in a required path can reduce the
end-to-end delay. A linear programming model is used to find the
shortest paths from $f$ flows in $F$:



\small
\begin{eqnarray}
  \textit{minimize } & & \left \{  \sum_{e_{sd} \in E }{a_{sd}} \right \} \label{eq:miniPath} \\
  \textit{Subject to}: \nonumber \\
  & & \sum_{d \in V}{a_{vd}} - \sum_{d \in V}{a_{dv}} = 1,\ \forall v \in V,\ v \neq i \label{eq:outEdges}\\
  & & \sum_{s \in V}{a_{si} = |F|},\ i = sink \label{eq:inSink} \\
  & & a_{sd} \geq 0 \label{eq:notNeg}
\end{eqnarray}
\normalsize


Variable $a_{sd}$ is defined as the sum of the flows using the edge
$e_{sd}$ on your route. Constraint (\ref{eq:outEdges}) ensures that
every vertex can only generate a single flow and constraint
(\ref{eq:inSink}) ensures that the sink node can receive a number of
flows equal to the number of sensors in the network. The combination
of restrictions (\ref{eq:outEdges}) and (\ref{eq:inSink}) ensures that
loops are not created. The constraint (\ref{eq:notNeg}) ensures
non-negative values for sum of flows on links in $a_{sd}$. The
objective function (\ref{eq:miniPath}) minimizes the sum of the amount
of all flows that pass through each edge $e_{sd} \in E$. Thus, each
flow is allocated to the shortest path and therefore the objective
function value is minimized.

\paragraph{Wireless Links with Low Quality} \label{pg:linkFracos}


The shortest paths may be more susceptible to high level of
interference that increases the packet loss rate, since the
interference decreases the link quality. The quality of the wireless
communication affects the overall network capacity. Wireless links
that do not have their quality degraded by factors, such as
interference and noise, become a good choice for use on routes to the
sink node, while links that have poor data transmission capacity
(i.e., high level of interference) should be avoided to improve
package delivery rate.


The LQI (Link Quality Indication) is a measurement provided by the
IEEE 802.15.4 physical layer that can be used as a quality metric of
the wireless transmission between two sensors. LQI values are
represented in a range from 0 to 255~\cite{Gomez2010}. In this paper,
a link is considered weak or with low quality if the LQI value is
smaller than a certain threshold ($TH_{LQI}$), according to the
following condition:
\begin{eqnarray}
  l_{sd} =
  \begin{cases}
    0 &,\  LQI_{sd}\ \geq \ TH_{LQI}\\
    1 - (\frac{LQI_{sd}}{TH_{LQI}})&,\  LQI_{sd}\ < \ TH_{LQI} \label{eq:normalizacao}\\
  \end{cases}
\end{eqnarray}


$LQI_{sd}$ is the value of LQI regarding the wireless link between
nodes $s,d \in V$. The variable $l_{sd}$ is used in the proposed model
to represent link quality values that should be avoided (higher ones)
or selected (lower ones). When $l_{sd}$ is zero, it indicates that the
link between the devices has an acceptable LQI, i.e. it has a good
quality. Otherwise, the value is normalized to the interval $(1,0)$ in
order to distinguish the links with lower quality, but that have a
good chance of being used, since they have a higher LQI.


The variable $l_{sd}$ is fundamental to minimize the use of links with
low quality on the paths. The objective function for this approach is
constructed combining the objective function (\ref{eq:miniPath}) and
taking into the number of hops (variable $l_{sd}$):
\begin{eqnarray}
  \textit{minimize } & & \left \{  \sum_{e_{sd} \in E }{ l_{sd} \cdot a_{sd}} \right \} \label{eq:miniLqi}
\end{eqnarray}

\paragraph{Network Bottleneck} \label{pg:gargalo}



The distribution of flows in a WSNs, in a balanced way, increases the
network lifetime. On the other hand, an unfair distribution of paths
for flows creates agglomeration flows on the same node
(i.e. bottleneck), increasing the power consumption due to the
increase of forwarded packets. Since the sensors usually have strong
restrictions relating to energy consumption, the question of
minimizing bottlenecks is a very important feature for WSN
scenarios. Furthermore, smoothing out network bottleneck can improve
traffic performance.


The variable $a_{sd}$ enables the objective function that is employed
to minimize the network bottleneck. Therefore, the objective function
is a combination of paths for each flow on the graph $G$, where the
highest sum of the amount of flows passing through a node $s$ is
minimal:

\begin{eqnarray}
  \textit{minimize } & & \left \{ max \left( \left \{ \sum_{d \in V}{a_{sd}} \right \}, \forall s \in V \right) \right \} \label{eq:miniBalanced}
\end{eqnarray}

\paragraph{Multi-objective Model}


The model of the multi-objective optimization problem for the routing
problem in WSNs uses three objectives (number of hops, quality links,
and network bottleneck):
\begin{eqnarray}
  \textit{minimize} & & \left \{  \sum_{e_{sd} \in E }{a_{sd}} \right \} \nonumber \\
  \textit{minimize } & & \left \{  \sum_{e_{sd} \in E }{ l_{sd} \cdot a_{sd}} \right \} \nonumber \\
  \textit{minimize } & & \left \{ max \left( \left \{ \sum_{d \in V}{a_{sd}} \right \}, \forall s \in V \right) \right \} \nonumber\\
  \text{Subject to:} \nonumber \\
  & & \sum_{d \in V}{a_{vd}} - \sum_{d \in V}{a_{dv}} = 1,\ \forall v \in V,\ v \neq i \nonumber \\
  & & \sum_{s \in V}{a_{si} = |F| - 1},\ i = sink \nonumber \\
  & & a_{sd} \geq 0 \nonumber 
\end{eqnarray}


The objective functions (\ref{eq:miniPath}), (\ref{eq:miniLqi}), and
(\ref{eq:miniBalanced}), when combined in a single model, create a
problem with conflicting solutions by having a set of optimal
solutions that are not dominated by each other, called the set of
Pareto optimal. In other words, the modeled objectives are conflicting
each other because there is no solution that is able to meet optimally
all three objectives.


The conflict between the objectives that minimizes the amount of hops
(\ref{eq:miniPath}), minimizes the amount of low-quality wireless
links (\ref{eq:miniLqi}), and minimizes the network bottleneck
(\ref{eq:miniBalanced}) is evident since the objectives
(\ref{eq:miniPath}) and (\ref{eq:miniLqi}) group the paths in order to
select shortest paths or a link with better quality, objective
function (\ref{eq:miniBalanced}) try to distribute the flows equally
over the paths on the entire network, therefore increasing the average
path length and the use of links with low quality. The objective
functions (\ref{eq:miniPath}) and (\ref{eq:miniLqi}) are conflicting
when the establishment of shortest route demands links with lower
quality.

\section{RALL --- Routing-Aware of path \\Length, Link quality, and \\traffic Load} \label{sec:descricaoAlgoritmo}


In this section we describe the multi-objective algorithm, called
\emph{Routing-Aware of path Length, Link quality, and traffic Load}
(RALL) for the routing problem in WSNs. RALL algorithm employs some
techniques to simplify the complexity of selecting the set of Pareto
optimal solutions. For this reason, it uses an algorithm for
determining paths with lower cost (Dijkstra's algorithm) so that each
path is calculated for a flow generated on an ordinary node. The
values related to the objective function are changed according to the
nodes that had already their specific route for the flows in order to
achieve the balancing. Initially, it is carried out the transformation
of the model so that the objective functions of path length
minimization (\ref{eq:miniPath}) and low quality links minimization
(\ref{eq:miniLqi}) are combined in a single objective.


When a solution of an objective function is not dominated for the
other objective, it is necessary to establish a trade-off between the
functions to determine which solution can satisfy partially or
completely the objectives. The weighted sum method of the objective
functions is usually applied to perform the combination of
objectives. Therefore, it is necessary to establish a weight for each
objective function and perform a weighted sum of these values. This
method was chosen because it is well known for the problems of linear
programming~\cite{Marler2009}:
\begin{eqnarray}
  minimize
  \left \{ w_p \cdot  \left( \sum_{e_{sd} \in E }{a_{sd}} \right)  +   w_l \cdot \left( \sum_{e_{sd} \in E }{ l_{sd} \cdot a_{sd}} \right) \right \} =  \nonumber 
\end{eqnarray}
\begin{eqnarray}
  \left \{ \sum_{e_{sd} \in E }{ \left( w_p + w_l \cdot l_{sd} \right) \cdot a_{sd} } \right \} \label{eq:combination}
\end{eqnarray}


$w_p\ and\ w_l$ are the weights for the objective function related to
the number of hops (\ref{eq:miniPath}) and number of low quality links
(\ref{eq:miniLqi}), respectively. As a requirement to apply this
approach, the decision variables of the objective functions must be at
the same scale or magnitude, it is not the case of the function
(\ref{eq:miniBalanced}). In the presented model, the decision variable
for the objectives is the number of flows passing through the links
(edges).


RALL algorithm begins by performing a combination of objective
functions path length minimization and low quality links minimization
in a single objective function, using the weighted sum
(\ref{eq:combination}). Next, RALL algorithm performs the minimization
of network bottleneck by updating the values of the objective
function. The algorithm~\ref{al:alagoritmo_principal} shows the RALL's
pseudo-code. It has a set of flows as input that are used to generate
the paths with lower cost.

\begin{algorithm}[!hbt] \label{al:alagoritmo_principal}
  \small\scriptsize
  \SetKwInOut{Input}{input}
  \SetKwInOut{Output}{output}
  \SetAlgoLined\DontPrintSemicolon	
  \SetKwFunction{RALL}{RALL}
  \SetKwFunction{SumWeights}{SumWeights}
  \SetKwFunction{UpdateEdges}{UpdateEdges}
  \SetKwProg{myalg}{Algorithm}{}{}
  \myalg{\RALL{}}{
    \Input{ \\
      $G -(V,E)$\\
      $O_f - \text{set of ordered flows}$ \\   		
      $i - \text{sink, } i \in V$ \\
      $w_p,w_l - \text{weights for objective functions}$\\}	
    
    \Begin {
   	$E \leftarrow $ \SumWeights($E$,$w_p$,$w_l$) \\
   	$A \leftarrow E$ \\
   	$P \leftarrow \emptyset$ \\
	\ForEach{$f_s \in O_f$} {
	  $p_s \leftarrow MCPath(s,\ i,\ G<V,A>)$ \\
	  $A \leftarrow $ \UpdateEdges{$p_s$, $E$, $A$}\\
	  $P \leftarrow P \cup \{p_s\}$
	}
      }
      \Return{$P$}
  }{}
  \SetKwProg{myprocSW}{Procedure}{}{}
  \myprocSW{\SumWeights{}}{
    \Input {
      \\$E - \textit{set of edges}$\\
      $w_p,w_l - \text{weights for objective functions}$ \\
    }
    \Begin {
   	$E_{new} \leftarrow \emptyset$ \\
	\ForEach{$e_{sd} \in E$} {
	  $e_{sd}^{new} \leftarrow p_{const} \cdot w_p + w_l \cdot l_{sd}$ \\
	  $E_{new} \leftarrow E_{new} \cup \{ e_{sd}^{new} \}$ \\
	}
	\Return{$E_{new}$}
      }
  }
  
  \SetKwProg{myprocAA}{Procedure}{}{}
  \myprocAA{\UpdateEdges{}} {
    \Input{
      \\$E - \textit{set of edges}$ \\
      $P_s - \text{router of  s to } sink$ \\
      \\$A - \textit{set of edges}$\\
    }	
    \Begin {
	\ForEach{$b_{jh} \in P_s$} {
	  $ r_{jh} \leftarrow  E \cap \{b_{jh}\} $ \\
	  $amount \leftarrow  \sum{a_{sd}}, \forall e_{sd} \in E, s = j $ \\
	  $ r_{jh} \leftarrow e_{jh} + amount, e_{jh} \in E $ \\
	  $ A \leftarrow A \cup \{ r_{jh} \} $
	}
	\Return{$A$}
      }
  }
  \caption{RALL (Routing-Aware of path Length, Link quality, and traffic Load.)}
\end{algorithm} 


On line 3, the weighted sum of objective functions is called. In this
procedure, path length and low quality links are combined in a single
vector cost. The variable $l_ {sd}$ is the normalized value of LQI for
edge $e_{sd}$, using the constraint shown by the equation
(\ref{eq:normalizacao}). $p_{const}$ is a constant used to establish
the value for each hop. The value of each link is updated according to
this constant value. If the chosen value for $p_{const}$ is small in
some interactions, the objective function to determine the lowest
number of hops is almost negligible, and the trade-off will be
impaired. This happens because the values that represent the
bottleneck in the objective function, on each iteration, are
incremented. It is used the value of $p_{const} = |V|$, and the values
of variable $l_{sd}$ are normalized to have $p_{const}$ as the
threshold value.


After the weighted sum is generated minimal cost path using
$MCPath(sender, \\ receiver, GRAPH)$ for each flow $f_s$, which is
assigned to the variable $p_s$ (on line 7). The links that make up the
minimum cost path $P_S$ will have their values updated by adding the
amount of flows that use the device $s \in V$ --- line 27 of
$UpdateEdges(p_s,E,A)$ procedure. Finally, the path is inserted into
the solution set $P$ that will contain all routes to every flow.


\section{Simulation Results} \label{sec:resultados}


We conducted a simulation study to analyze the performance of proposed
approach. A network with a single sink node and many sensor nodes are
taken into consideration in the simulated network. The sink node
receives data from the sensor nodes and it serves as a connection
point to an external network. The sensor nodes send the collected data
and forward/relay packets on a single route to the sink node. All
nodes are static and they use asymmetric links. This section is
organized as follows: the scenario configuration is outlined in
sub-section \ref{scenarioConfiguration}. The evaluation and the
simulation results are discussed in sub-section \ref{subResults}.

\hyphenation{OMNeT}


\subsection{Scenario Configuration} \label{scenarioConfiguration}

We employ the Castalia module~\cite{Boulis2011} of OMNeT++
simulator~\cite{Varga2001}, which are widely used to evaluate
WSN. Table~\ref{tab:cenario} shows the general parameters used in the
simulation and Table~\ref {tab:motes} presents node configuration
based on~\cite{Moghadam2014}.



The simulations were performed in scenarios varying the number of
nodes, i.e. 10 up to 50 nodes (Table~\ref{tab:cenario}). For every set
of nodes, it was generated ten topologies and four random seeds,
resulting in a total of 40 runs for each set of nodes. The positions
of the nodes in a topologies were generated using a normal
distribution, limited in area, that guarantees none of the nodes would
remain isolated from the network (out of range from any other node).

\begin{table}[!hbt]
  \centering
  \caption{Simulation Parameters}
  \label{tab:cenario}
  \begin{tabular}{|l|l|}
    \hline
    Packet Generation Rates  & 5 packets/minute                \\ \hline
    Traffic generation model & Poisson                         \\ \hline
    Topology Area            & 50x50                           \\ \hline
    Number of nodes          & 10, 20, 30, 40, and 50          \\ \hline
    Interference Model       & Additive \footnotemark[1]       \\ \hline
    \parbox[t]{4cm}{Weight values for path \\length function ($w_p$)}       & 50\% \\ \hline
    \parbox[t]{4cm}{Weight values for link \\quality function ($w_l$)}  & 50\%     \\ \hline
  \end{tabular}
\end{table}

  
\footnotetext[1]{Additive interference: it use the SINR values of the
  package, the receiver receives the strongest signal of the two
  transmissions (if it is strong enough).}

\begin{table}[!hbt]
\centering
\caption{Node Configuration}
\label{tab:motes}
\begin{tabular}{|l|l|}
\hline
Initial Energy                       & 100 J                     \\ \hline
TX Energy                            & 20 mJ/pkt                 \\ \hline
RX Energy                            & 10 mJ/pkt                 \\ \hline
MAC layer                            & T-MAC                     \\ \hline
Radio Model                          & CC2420                    \\ \hline
Transmission Power                   & -10dBm                    \\ \hline
Packet Size	                     & 127 bytes                 \\ \hline
Link Communication                   & Asymmetric                \\ \hline
\end{tabular}
\end{table}


The simulation was carried out in two phases. In the first phase, the
neighborhood discovery was done by disseminating broadcast
messages. These control messages are also used to calculate the
average LQI value of the link for each neighbor so that the sink node
has the complete network graph information. Due to the fact that the
processing capacity and energy of the sensors devices are very
restricted, the routing algorithm is performed in a centralized way in
the sink node.


Once in possession of the network global view, the proposed routing
algorithm is executed to select the path to the sink for each sensor
node. The second simulation phase uses the routing table information
to forward the packets. Each node sends messages to the sink, which
will extract some of the performance metrics for analysis.

\subsection{Results} \label{subResults}


Other routing algorithms were chosen to assist the performance
assessment of RALL, as following:

\begin{itemize}
  

\item $BALANCED-LQI$: It is based on similar criteria of RALL,
  however, it only employs the load balancing and link quality
  objective functions, and it does not perform the combination of
  functions using the weighted sum.

  
\item $BPR$: It is a heuristic with two objective functions, load
  balancing and path length~\cite{Mello14}. For each flow generated on
  the network, BPR seeks to select the shorter route belonging to set
  of candidate paths that does not increase the network bottleneck.

\item $PATH$: Traditional approach to determine the shortest path
  length (e.g. Dijkstra's algorithm).


\item $LQI$: It aims to select routes that use fewer links with low
  quality. A link is considered low quality when it does not satisfied
  the described condition in (\ref{eq:normalizacao}).
  
\end{itemize}


The comparison of these approaches aims to provide better impact of
the different criteria in the traffic and network performance through
the route selection.

\begin{figure}[!hbt]
  \centering
  \centering
  \includegraphics[width=1.0\textwidth]{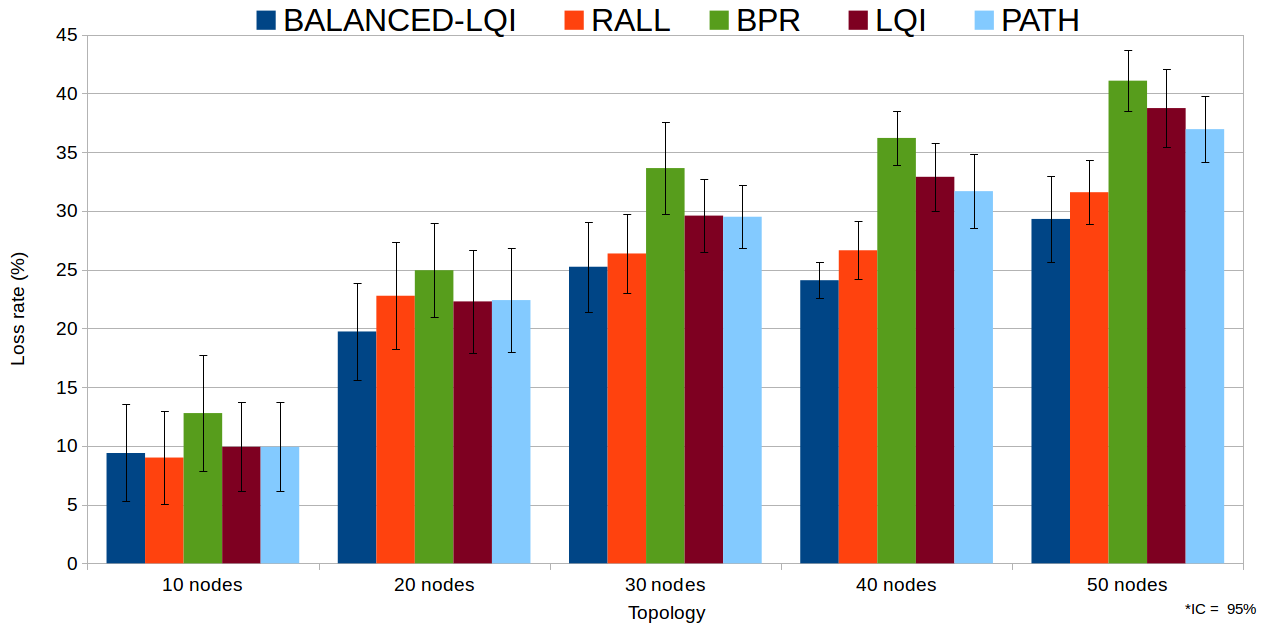}
  \caption{Packet loss rate.}
  \label{fig:perdaPacotes}
\end{figure}
\begin{figure}[!hbt]
  \centering
  \includegraphics[width=1.0\textwidth]{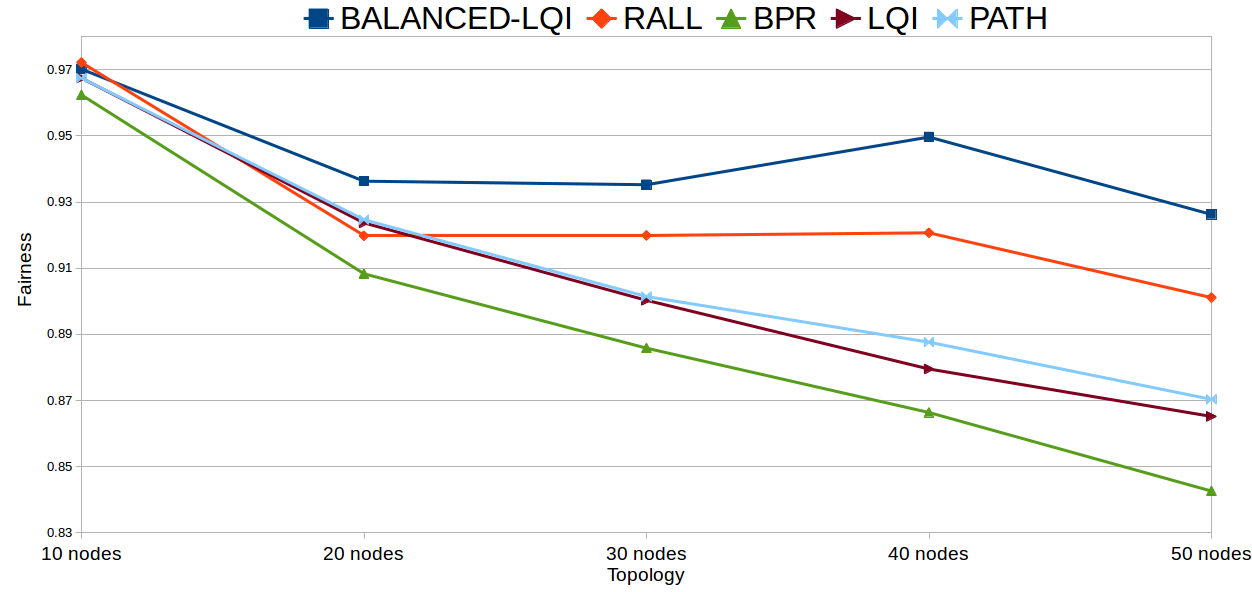}
  \caption{Jain fairness index.}
  \label{fig:justica}
\end{figure}

Figure~\ref{fig:perdaPacotes} shows the packet loss rate. We figure
out that the $BALANCED-LQI$ and $RALL$ algorithms achieve lower
levels of losses, especially in high density network (i.e. 50
nodes). Therefore, the amount of data increases. This can be justified
by the Jain fairness index (well-known metric to evaluate load
balancing) shown in Figure~\ref{fig:justica}, and by the fact that the
two algorithms take into consideration the link quality. Despite the
fact that $BALANCED-LQI$ slightly results in lower packet loss, the
confidence interval (Figure~\ref{fig:perdaPacotes}) shows that the
performance of both approaches is very similar.



%

\begin{figure*}[!hbt]
  \centering
  \subfigure[10 nodes]{\includegraphics[width=1.00\textwidth]{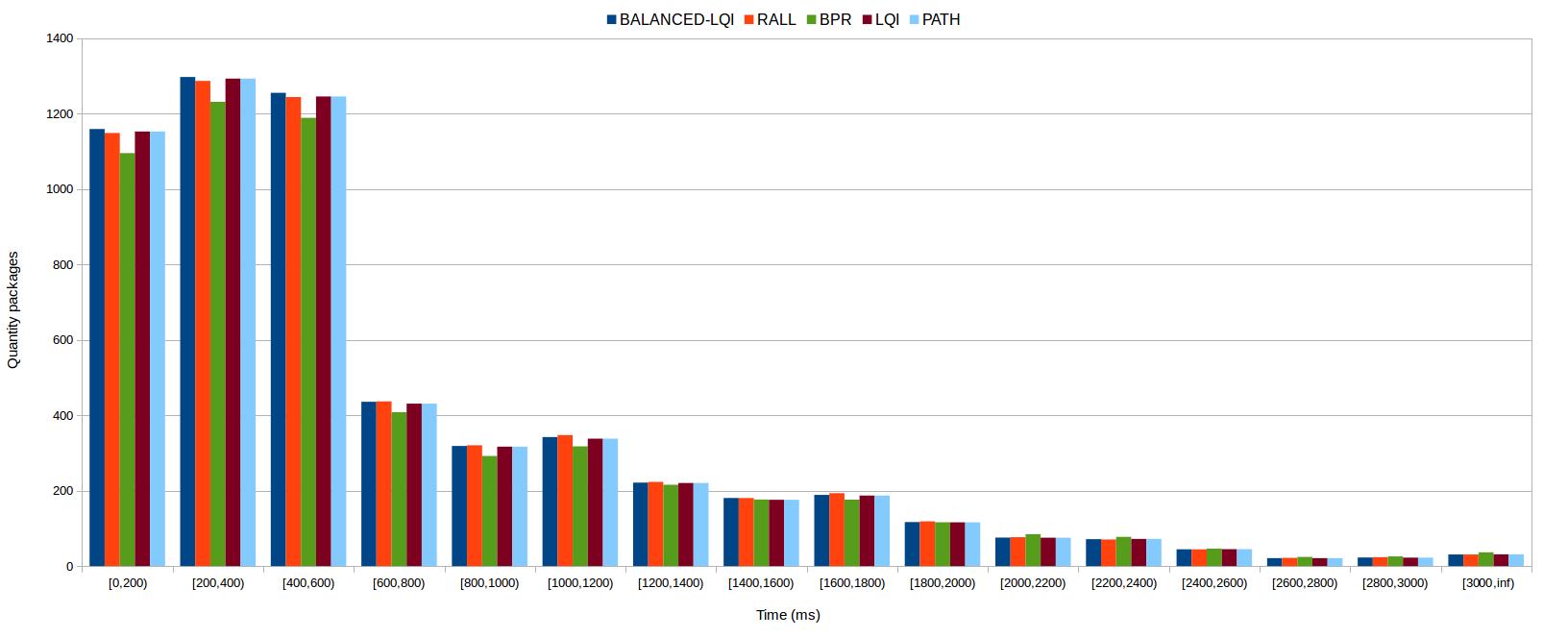}} \label{fig:latencia10}
  \subfigure[20 nodes]{\includegraphics[width=0.20\textwidth]{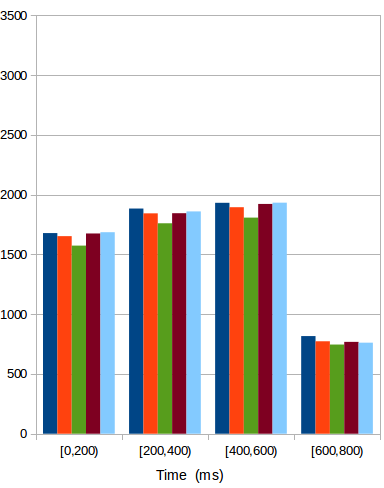}} \label{fig:latencia20}
  \subfigure[30 nodes]{\includegraphics[width=0.20\textwidth]{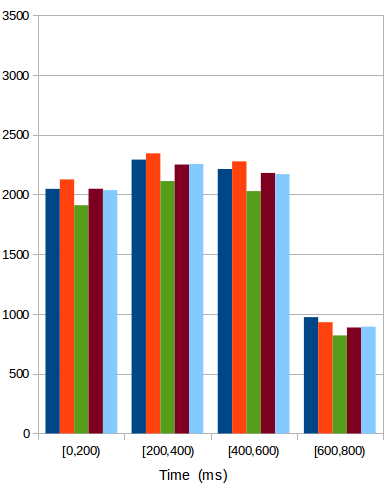}} \label{fig:latencia30}
  \subfigure[40 nodes]{\includegraphics[width=0.20\textwidth]{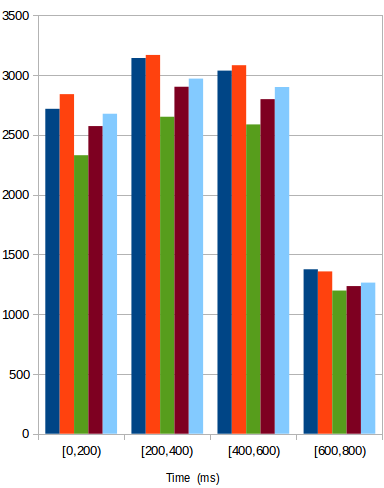}} \label{fig:latencia40}
  \subfigure[50 nodes]{\includegraphics[width=0.20\textwidth]{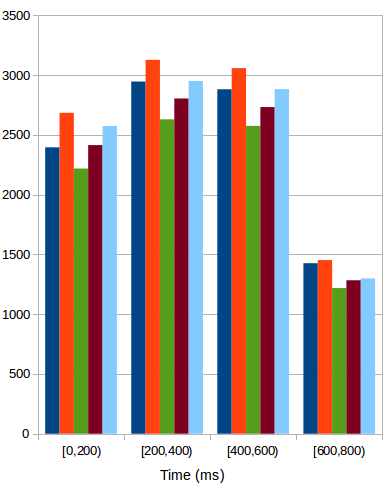}} \label{fig:latencia50}
  \caption{Latency for 10, 20, 30, 40, and 50 nodes.} \label{fig:latencia}
\end{figure*}

Figure~\ref{fig:latencia} shows the average latency in data delivery
of the simulated network scenarios. It is worth noting that in the
scenario with 10 nodes, most of packets were delivered within the
range of $[0,600)$ms. The other intervals follow the behavior of the
first three ones. In fact, it occurs in all scenarios, thus we show
only the first four intervals: 20, 30, 40, and 50.

\begin{figure}[!hbt]
\centering
\includegraphics[width=1.00\textwidth]{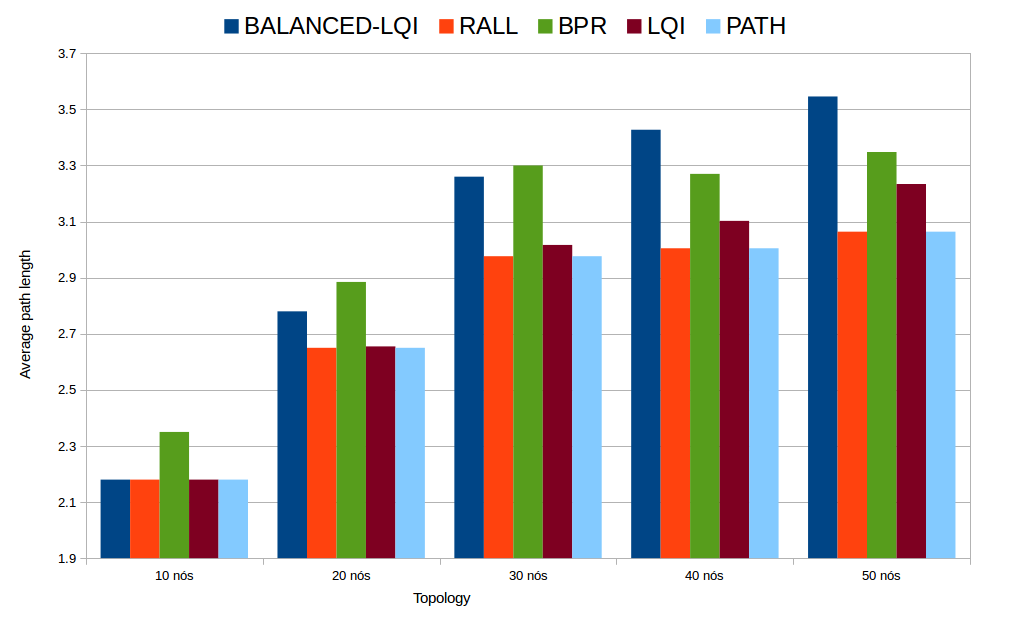}
\caption{Average path length.}
\label{fig:avgLenPath}
\end{figure}

The greatest difference in latency between approaches can be noticed
when the number of nodes increases. Again, the $BALANCED-LQI$ and
$RALL$ approaches are featured, however, $RALL$ results in an
significant improvement of the latency, followed by $PATH$. These
results can be justified by analyzing the average path length
generated by each approach (Figure~\ref{fig:avgLenPath}), where $RALL$
and $PATH$ reached the lowest values.

\begin{figure}[!hbt]
  \centering
  \includegraphics[width=0.8\textwidth]{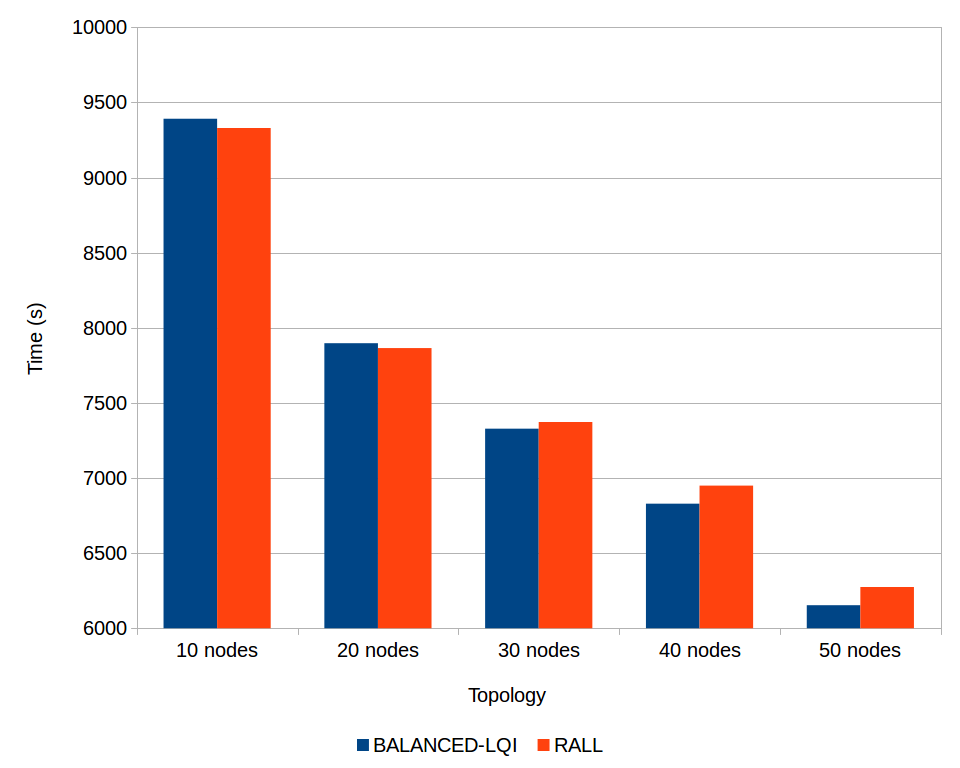}
  \caption{Average network lifetime.}
  \label{fig:tempoVida}
\end{figure}

We also analyzed the average network lifetime, which was measured as
from the beginning of the second phase up to the first node of the
network run out the battery energy. In this analysis we used only
$BALANCED-LQI$ and $RALL$ approaches, because they showed better
performance for packet loss and latency. Figure~\ref{fig:tempoVida}
illustrates the lifetime.


\begin{figure}[!hbt]
  \centering
  \includegraphics[width=0.8\textwidth]{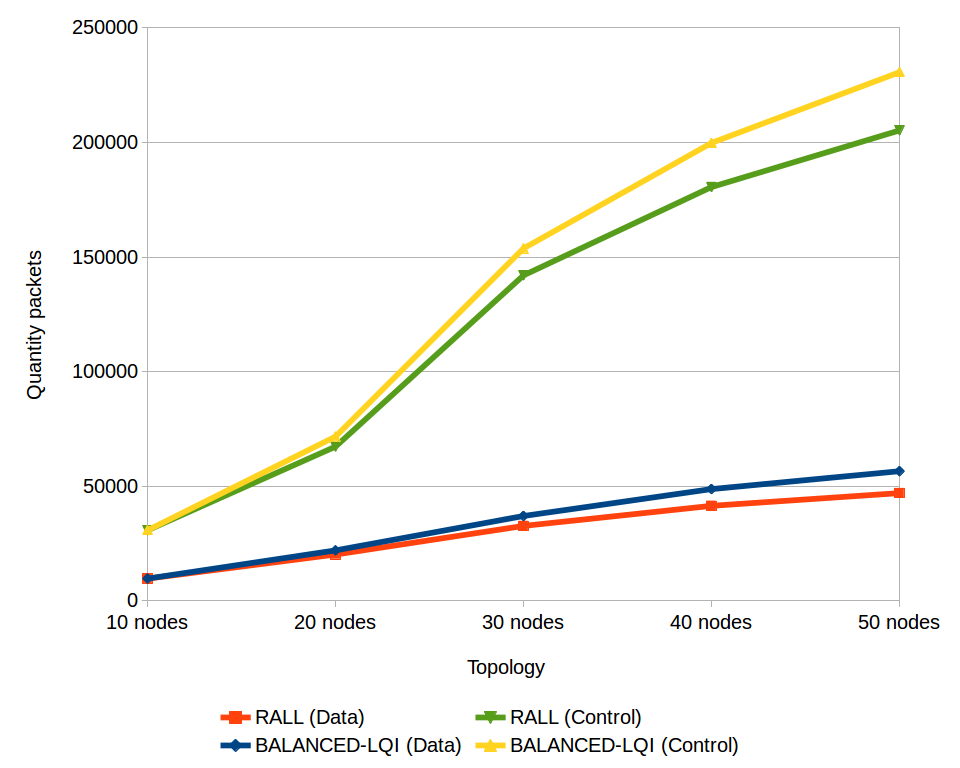}
  \caption{Average control overhead in MAC layer.}
  \label{fig:controleVsDados}
\end{figure}


$RALL$ approach achieves performance improvement when increasing the
number of nodes (i.e. 30, 40, and 50 nodes). In order to better
understand this behavior, we analyzed the packets exchanged by each
algorithm. In Figure~\ref{fig:controleVsDados}, data packets are
generated by the application layer and accounted in packet loss by the
sink. Control packets are accounted of the MAC layer, which uses the
T-MAC protocol. This protocol employs an adaptive method to avoid
collision and to control the radio duty-cycle, in order to reduce the
energy consumption.


It is important to stress out that $BALANCED-LQI$ has a slightly
larger number of data packets delivery to all scenarios. Moreover,
$BALANCED-LQI$ also requires a larger amount of control packets in the
MAC layer. Each packet has an energy cost for sending and
receiving. This explains why the network lifetime achieves the highest
times by getting shortest paths (activating fewer links).

\section{Related Work} \label{sec:relatedwork}


There are many works that seek to combine these objectives, some of
them to optimize routing in wireless networks, especially in WSNs. The
most of these works deal with one or two objectives described
above. Wang and Zhang~\cite{Wang2010} presented a protocol, called
Interference Aware Multipath Routing (IAMR), that selects paths
spatially disjoint, adopting an interference model in which
interference range is twice greater than the transmission range. This
model may overestimate the interference level.


Radi \emph{et al.}~\cite{Radi2010} proposed Low-Interference
Energy-Efficient Multipath Routing Protocol (LIEMRO), which takes into
account a load balancing multipath approach and which also seeks to
minimize interference through the use of ETX routing metric. As it was
described in the literature, ETX has limitations to capture the
interference in a more accurate way~\cite{Borges11}.


Minhas \emph{et al.}~\cite{Minhas2009} recommended an approach based
on fuzzy logic for multi-objective optimization problem in routing
that seeks to balance two objectives: to maximize the network lifetime
(analyzing the current and residual energy levels of the nodes) and
the delay between the ordinary node and the sink node (taking into
account of the number of hops).


Load balancing routing can provide both packet loss minimization and
network lifetime maximization based on a uniform distribution of flows
in the network. For this reason, routing approaches are proposed in
wireless networks to minimize bottlenecks and the average path length,
such as Bottleneck, Path Length and Routing heuristic overhead
(BPR)~\cite{Mello14}.


Moghadam \emph{et al.}~\cite{Moghadam2014} presented a heuristic,
named Heuristic Load Distribution (HeLD), which aims to minimize the
overhead communication and maximize the network lifetime by a load
balancing routing based on a linear programming approach, and the use
of a set of braided paths. Simulation results showed that HeLD
achieved less overhead and an increase in network lifetime, however,
packet delivery rate was lower and latency was not presented. This can
be partially justified by braided paths that can generate a high level
of interference.


Machado \emph{et al.}~\cite{Machado2013} presented a heuristic for IoT
environments called Routing by Energy and Link quality (REL), which
seeks to select paths that minimize the number of hops, maximize
network lifetime through the residual energy of the nodes, and
maximize the link quality of the in the selected paths based on
Received Signal Strength Indicator (RSSI) metric. However, REL is
quite simple and it is limited to establish a balance among the three
objectives.


None of the related work handles the three mentioned objectives. Most
of them use heuristics for routing solution. This paper propose a new
multi-objective routing algorithm for WSNs. The algorithm seeks to
achieve the trade-off among load balancing, path length, and
interference, i.e. aggregate flows in the shortest paths to avoid
overloaded links and with high level of interference.

\section{Conclusions and Future Work} \label{sec:conclusao}


We propose a multi-objective routing algorithm (RALL) for IoT
environments in this paper. RALL seeks to minimize three objectives:
bottlenecks, links with high levels of interference, and long paths.


The results showed that RALL algorithm results in good levels of
packet loss rate (very similar to the main comparative approach), the
lowest latency, and longer average network lifetime, mainly in dense
scenarios.


Although RALL does not reach the best results in all tests, it shows
the importance of the all employed criteria in order to offer an
overall routing approach for IoT. The weakness of one criterion is
smoothed out by the others as well as the equilibrium of the criteria
to provide a more efficient performance for WSNs. We believe that the
proposed algorithm is an interesting approach to IoT environments that
have different types of traffic and require routing approaches that
seeks to balance conflicting objectives to improve delivery rate,
delay, and power consumption.


As future work, we intend to start further studies about the weights
used in the objective functions, for example, trying to make a best
correlation between the weights and types of topologies. Moreover, the
RALL algorithm will be implemented in testbeds for performing
experiments in a real environment.

\bibliographystyle{plain}
\bibliography{arxiv-0001}

\end{document}